\documentclass[twocolumn]{aastex631}

\usepackage{graphicx}

\newcommand{\mum}{\ifmmode{\rm \mu m}\else{$\mu$m}\fi}

\shorttitle{The evolved stellar population of Sextans A}
\shortauthors{Gavetti et al.}

\begin{document}

\title{Dust in the very metal-poor galaxy Sextans A with JWST. I: Characterizing the evolved stellar population of Sextans A based on JWST observations and stellar evolution models}

\author{C.\ Gavetti}
\affiliation{Dipartimento di Matematica e Fisica, Università degli Studi Roma Tre, Via della Vasca Navale 84, 00100, Roma, Italy}
\affiliation{INAF - Osservatorio Astronomico di Roma, Via Frascati 33, 00078, Monteporzio Catone, Roma, Italy}

\author[0000-0003-2442-6981]{F.\ Dell'Agli}
\affiliation{INAF - Osservatorio Astronomico di Roma, Via Frascati 33, 00078, Monteporzio Catone, Roma, Italy}

\author[0000-0003-1356-1096]{E.\ Tarantino}
\affiliation{Space Telescope Science Institute, 3700 San Martin Drive, 
Baltimore, MD 21218, USA}

\author[0000-0003-4850-9589]{M. L.\ Boyer}
\affiliation{Space Telescope Science Institute, 3700 San Martin Drive, 
Baltimore, MD 21218, USA}

\author[0000-0003-0356-0655]{I.\ McDonald}
\affiliation{Jodrell Bank Centre for Astrophysics, Alan Turing Building, University of Manchester, Manchester M13 9PL, UK; Open University, Walton Hall, Kents Hill, Milton Keynes MK7 6AA, UK}

\author[0000-0002-1272-3017]{J.~Th.\ van Loon}
\affiliation{Lennard-Jones Laboratories, School of Chemical \& Physical Sciences, Keele University, ST5 5BG, UK}

\author[0000-0002-1693-2721]{D.~A.\ Garc{\'\i}a-Hern{\'a}ndez}
\affiliation{Instituto de Astrof{\'\i}sica de Canarias, E-38205 La Laguna, Tenerife, Spain}
\affiliation{Departamento de Astrof{\'\i}sica, Universidad de La Laguna, E-38206 La Laguna, Tenerife, Spain}

\author[0000-0003-2723-6075]{M.A.T.\ Groenewegen}
\affiliation{Koninklijke Sterrenwacht van Belgi\"e, Ringlaan 3, B--1180 Brussels, Belgium}

\author[0000-0001-6652-1069]{A.\ Nanni}
\affiliation{National Centre for Nuclear Research, ul. Pasteura 7, 02-093 Warsaw, Poland}
\affiliation{INAF - Osservatorio astronomico d'Abruzzo, Via Maggini SNC, 64100, Teramo, Italy}

\author[0000-0002-5797-2439]{J.A.D.L.\ Blommaert}
\affiliation{Astronomy and Astrophysics Research Group, Department of Physics and Astrophysics, Vrije Universiteit Brussel, Pleinlaan 2, B-1050 Brussels, Belgium}

\author[0000-0003-1319-4089]{R. D.\ Gehrz}
\affiliation{Minnesota Institute for Astrophysics, School of Physics and Astronomy, 116 Church Street SE, University of Minnesota, Minneapolis, MN 55455, USA}

\author{L. M.~Gerlach}
\affiliation{Lennard-Jones Laboratories, School of Chemical \& Physical Sciences, Keele University, ST5 5BG, UK}

\author[0000-0002-8937-3844]{S.\ Goldman}
\affiliation{Space Telescope Science Institute, 3700 San Martin Drive, 
Baltimore, MD 21218, USA}

\author[0000-0001-9910-9230]{M. Marengo}
\affiliation{Department of Physics, Florida State University, Tallahassee, FL 32303, USA}

\author[0000-0001-5538-2614]{K.B.W. McQuinn}
\affiliation{Space Telescope Science Institute, 3700 San Martin Drive, Baltimore, MD 21218, USA}
\affiliation{Department of Physics and Astronomy, Rutgers, The State University of New Jersey, 136 Frelinghuysen Rd, Piscataway, NJ 08854, USA}

\author[0000-0002-0861-7094]{J. M. Oliveira}
\affiliation{Lennard-Jones Laboratories, School of Chemical \& Physical Sciences, Keele University, ST5 5BG, UK}

\author[0000-0001-6326-7069]{J.\ Roman-Duval}
\affiliation{Space Telescope Science Institute, 3700 San Martin Drive, Baltimore, MD 21218, USA}

\author[0000-0002-6858-5063]{R.\ Sahai}
\affiliation{Jet Propulsion Laboratory, California Institute of Technology, Pasadena, CA 91109, USA}

\author[0000-0003-0605-8732]{E. D.\ Skillman}
\affiliation{Minnesota Institute for Astrophysics, School of Physics and Astronomy, 116 Church Street SE, University of Minnesota, Minneapolis, MN 55455, USA}

\author[0000-0002-6858-5063]{B. F.\ Williams}
\affiliation{Astronomy Department, University of Washington, Seattle, WA 98195, USA}

\author[0000-0001-8392-6754]{A.\ Javadi}
\affiliation{School of Astronomy, Institute for Research in Fundamental Sciences (IPM), Tehran, 19568-36613, Iran}

\author[0000-0003-4870-5547]{O.\ C.\ Jones}
\affil{UK Astronomy Technology Centre, Royal Observatory, Blackford Hill, Edinburgh, EH9 3HJ, UK}

\author[0000-0003-2743-8240]{F.\ Kemper}
\affil{Institut de Ciències de l'Espai (ICE, CSIC), Can Magrans, s/n, E-08193 Cerd
anyola del Vallès, Barcelona, Spain}
\affil{ICREA, Pg. Lluís Companys 23, E-08010 Barcelona, Spain}
\affil{Institut d'Estudis Espacials de Catalunya (IEEC), E-08860 Castelldefels, Barcelona, Spain}

\author{F.\ La Franca}
\affiliation{Dipartimento di Matematica e Fisica, Università degli Studi Roma Tre, Via della Vasca Navale 84, 00100, Roma, Italy}

\author[0000-0003-4520-1044]{G.~C.\ Sloan}
\affiliation{Space Telescope Science Institute, 3700 San Martin Drive, 
Baltimore, MD 21218, USA}
\affiliation{Department of Physics and Astronomy, University of North Carolina, 
Chapel Hill, NC 27599-3255, USA}

\begin{abstract}
The nearby star-forming dwarf galaxy Sextans A offers a unique window into galaxy evolution in the early Universe, owing to its extremely low metallicity ($\sim$1--7 \% Z$_\odot$). Recent JWST imaging of Sextans A spanning 1--21~$\mu$m enables a detailed characterization of its dusty stellar populations and interstellar medium. In this work, we compare the observed JWST color-magnitude distributions of evolved stars with stellar evolution and dust-formation models to characterize the properties of the asymptotic giant branch (AGB) population, including progenitor mass, formation epoch, metallicity and dust production. Evolutionary tracks for $0.8$--$7$~M$_\odot$ stars with metallicity Z=$10^{-3}$ provide good agreement with the overall distribution of AGB stars in Sextans A. More than 90\% of the AGB population occupies a nearly vertical sequence in the color-magnitude diagrams, corresponding to stars spanning a wide range of masses and ages but exhibiting little or no circumstellar dust. This sequence appears to be dominated by oxygen-rich (M-type) AGB stars and reveals that the F444W flux is a robust luminosity diagnostic. A small subset of sources displays strong infrared excesses and is dominated by carbon stars descending from $1.25$--$1.5$~M$_\odot$ progenitors that formed $\sim$2--3~Gyr ago and are currently in the final AGB phases. Their MIRI colors imply very low metallicities, consistent with estimates from the red giant branch morphology ($\sim$1--2 \% Z$_\odot$). Finally, we show that the JWST/NIRCam F277W-F444W color serves as an effective proxy for the dust production rate, with models predicting rates up to $\sim 10^{-7}$~M$_\odot/{\rm yr}$ for the reddest sources in Sextans A.

\end{abstract}


\section{Introduction}
\label{s.intro} 

The asymptotic giant branch (AGB) is the evolutionary phase experienced
by all the stars of mass below $\rm \sim 8~M_{\odot}$, which follows core helium-burning and precedes the white dwarf cooling.
During the AGB phase the stars are supported by a CNO-burning shell, with 
periodic ignitions of a helium layer lying above the degenerate core, 
which are commonly referred to as thermal pulses, because helium-burning occurs in thermally unstable conditions \citep{schw}. 
The surface chemical
composition of AGB stars can be altered by two physical mechanisms,
namely third dredge-up and hot bottom burning.
The former consists of the inward penetration of the surface
convection to regions of the star touched by helium-burning
nucleosynthesis, thus significantly enriched in carbon \citep{iben}: a
series of third dredge-up events causes the surface carbon abundance to exceed that of oxygen, and
hence the formation of carbon stars. Hot bottom burning consists of the
activation of advanced proton-capture nucleosynthesis at the 
base of the convective envelope, once the temperatures in those
regions reach and exceed $3 \times 10^7$ K \citep{sack}: following the ignition
of hot bottom burning the surface chemistry is altered on the basis of the
equilibria of the proton-capture nucleosynthesis experienced, which
is to be extremely sensitive to the metallicity of the stars
\citep{flavia18}.

Several studies over the past decades have convincingly shown that stars evolving through the AGB are highly efficient dust manufacturers. This efficiency arises from the physical conditions in their extended circumstellar envelopes, which are particularly favorable for the condensation of gaseous molecules into solid particles \citep{gail85, gail99, hofner18}. What remains unclear is the relative contribution of AGB stars with respect to supernovae (SNe) to the overall dust budget of the Universe (Schneider \& Maiolino 2024), an issue further complicated by the uncertainties affecting the description of dust production during the SN explosion (Bianchi \& Schneider 2007; Todini \& Ferrara 2001; Bocchio et al. 2016). 
Assessing the relative importance of AGB stars and SNe as sources of dust and mass in different environments and cosmic epochs is a complex task, considering that
a significant, though largely unknown, fraction of the dust formed in SNe may be destroyed by the reverse shock \citep{laki15}, and that the dust produced by either class of stars is exposed to destruction through subsequent interstellar medium processes.

To quantify the dust produced by low- and intermediate-mass stars during the AGB phase, some research teams combined models of AGB evolution with those of dust formation in the circumstellar envelope \citep{ventura12, ventura14, nanni13, nanni14}. These models are based on the description of a static wind expanding from the stellar surface, as proposed by \citet{fg02, fg06}. 
These developments made it possible to determine the 
mineralogy and the quantity of dust produced by the stars during different
evolutionary phases, to predict the time evolution of the spectral energy 
distribution (SED) and to estimate the dust yields for the different species.

The modelling described above can be applied to single stars only, which is the focus of the present investigation. We believe it is important to
stress that binary interactions may also 
influence the dust properties of evolved low- and intermediate-mass 
stars \citep{decin21}. Indeed, episodes of envelope stripping or common-envelope evolution can 
significantly modify the mass-loss history and the chemical composition 
of the material expelled \citep{flavia21}, potentially producing sources with dust 
characteristics that deviate from those expected from isolated AGB 
evolution. Examples of the effects of binary interaction 
include some post-RGB and post-AGB objects, which are 
known to display substantial mid-infrared excess and dusty circumstellar 
structures \citep{sahai07, sahai22}. While a detailed treatment of binary channels 
lies beyond the scope of this work, their possible contribution should 
be kept in mind when interpreting the dust-rich component of evolved 
stellar populations.

A valuable opportunity to validate and improve the theoretical description of the dust
production mechanism in the wind of AGB stars is offered by the study of the evolved stellar populations of galaxies. The formation of dust grains modifies the SED of the stars, thus affecting the morphology
of the evolutionary tracks in the various observational color-magnitude diagrams. This
can be compared with the observed distribution of the stars in the same
diagrams. Early attempts to interpret the AGB population of galaxies by means
of results from stellar evolution modelling in which the formation
of dust in the circumstellar envelope is considered, mostly based on data 
collected from the Hubble Space Telescope, Spitzer and 2MASS, were first applied 
to the Magellanic Clouds \citep{flavia14, flavia15a, flavia15b, nanni18, nanni19}, 
to a few Local Group galaxies \citep{flavia16, flavia18}, and more recently to 
M31 \citep{cla25}. In these studies part of the effort was devoted to
relating the distribution of the sources in the various regions of the
observational diagrams considered with the star formation history (SFH) and the
age--metallicity relation of the galaxies investigated.

The advent of the James Webb Space Telescope \citep[JWST; ][]{gardner23} provides substantial support to this line of research, because it opens the way to investigating the evolved stellar populations of all the Local Group galaxies, and possibly beyond \citep{nally24}.
JWST data hold promise to improve upon previous state-of-the-art programmes such as DUSTiNGS \citep{boyer15a, boyer15b, 
boyer17, goldman19}, the Sagittarius dSph \citep{mcd13, mcd14}, and M32 \citep{jones15, jones23}.
The interpretation of the incoming JWST observations will therefore be crucial for making a decisive step toward a comprehensive understanding of dust formation mechanisms in the circumstellar envelopes of evolved stars and, more generally, for assessing their contribution to the dust budget of their host systems and of the Universe as a whole.

We compare state-of-the-art dust formation and stellar evolutionary models to new JWST observations of the dwarf galaxy Sextans A. This 
galaxy is particularly interesting for studying the process of dust formation, because it 
harbors a metal-poor population of stars with metallicity between $1\%$ and $7\%$ of 
solar \citep{kniazev05, mcc12, sakai96}. Consequently, it can provide important insights into the metallicity dependence of dust production, a topic for which no general consensus has yet been reached \citep{vanloon08, sloan12, sloan16}. 
The data were observed as part of program PID: 1619 (PI M. Boyer; \citet{boyer25}, E. Tarantino et al. 2026, in preparation), and the combination of NIRCam \citep{rieke23} and MIRI \citep{wright23} data enables an exhaustive analysis of the entire sample of AGB stars. Tarantino et al. (2026, in preparation) present the JWST stellar catalog, including the classification of the dusty stellar populations and color-based estimates of the dust production for individual stars. Henry et al. (2026, in preparation) present full SED fitting of the evolved stars from 1--21~$\mu$m. In this paper, we focus on characterizing the evolved star population of Sextans A by comparing the data to stellar evolution models to constrain progenitor masses, formation epochs, metallicities, and dust production. Particular attention is given to the six stars for which MIRI (5--12~$\mu$m) low-resolution spectroscopy (LRS) is available \citep{boyer25}. We also aim to improve the theoretical framework for dust formation.

The paper is structured as follows: section \ref{input}
describes the set of evolutionary tracks adopted in the present
work and the schematization adopted to model dust formation in
the wind and the evolution of the SED; the path of the evolutionary tracks of stars of different mass on the 
color-magnitude diagrams built with the NIRCam filters, and
a gross distinction between dust-free objects of Sextans A
and red, dusty stars, is discussed in section \ref{nircam};
section \ref{miri} is focused on the interpretation of the
MIRI photometry; the discussion on the dust produced by the
individual sources is presented in section \ref{dpr}; finally, 
the conclusions are given in section \ref{concl}.

\section{Physical input and data}
\label{input}
To study the evolved stellar population of Sextans A, we use observations from the JWST-GO-1619 program, where imaging of most of Sextans A’s star-forming disk was obtained. Both NIRCam and MIRI instruments were used to map the galaxy in the F090W, F140M, F150W, F200W, F277W, F300M, F335M, F444W, F770W, F1000W, F1130W and F1280W bands. Preliminary Vega-based magnitudes from NIRCam and MIRI\footnote{Data were reduced using pipeline version 1.15.1 and the CRDS context jwst$\_$1293.pmap.} were measured using DOLPHOT, the point-spread-function photometry package developed by \citet{dolphin00, dolphin16} and adapted for JWST by the JWST Stellar Populations Early Release Science team \citep{weisz23, weisz24}. Here we briefly describe the procedure followed, while the detailed description of the imaging and photometry of the galaxy will be discussed in a forthcoming paper (E. Tarantino et al. 2026, in preparation). We followed the photometric quality cuts recommended by \citet{weisz23} and \citet{warfield23}, which are designed to separate stars and unresolved background galaxies, minimizing background contaminates. The field of view consists of a 2$\times$1 MIRI mosaic, covering the bulk of the disk of Sextans~A, and is 2.5 arcmin by 3.75 arcmin in size. The photometric catalog is complete up to 25 magnitudes in all NIRCam filters, much fainter than the magnitudes of the bright AGB stars studied in this work. We adopt photometric uncertainties that include the photon-noise characteristics reported by DOLPHOT, of $<0.005$ mag for NIRCam bands and $<0.05$ mag for MIRI bands in the magnitude regime analyzed. The final uncertainties tend to be 3–10$\times$ higher in HST data with a similar depth and crowding \citep{Williams14}. The JWST data presented in this article were obtained from the Mikulski Archive for Space Telescopes (MAST) at the Space Telescope Science Institute. The specific observations analyzed can be accessed via \dataset[doi:10.17909/edr6-ny09]{https://doi.org/10.17909/edr6-ny09}.

We interpret the distribution of the stars on the observational color–magnitude diagrams using evolutionary tracks for different stellar masses at metallicity Z = 0.001 and $\alpha$-enhancement $\rm [\alpha/Fe]=+0.4$ (corresponding to $\rm [Fe/H] \sim -1.4$), which are appropriate for the low-metallicity stellar population of Sextans A \citep{mcc12, sakai96}. We adopt the evolutionary sequences at Z = 0.001 from \citet{ventura14}, further refined and extended to the post-AGB and planetary nebula phases by \citet{devika23}.
 
Starting from the results of stellar evolution modelling, 
the calculation of the evolutionary tracks was based on
a three-step process, consisting of: a) the modelling of the dust 
formation process from the start of the core helium-burning phase to the beginning
of the post-AGB evolution; b) the determination of the synthetic SED; c) the calculation of the
magnitudes in the various NIRCam and MIRI filters. This procedure follows the methodology introduced
by \citet{flavia14, flavia15a}, and which was applied to 
Spitzer photometric observations of the Magellanic Clouds, in order
to interpret their evolved stellar population.

For some specific points along the AGB phase (typically ten points equally
spaced in time during each interpulse phase), we simulated the dust formation 
process in the circumstellar envelope based on the values of the luminosity,
effective temperature, mass-loss rate and surface chemical composition
of the star, predicted by the stellar evolution modelling. This step allows the determination of the quantity and the mineralogy 
of the dust formed in the stellar wind, the dust production rate (DPR) and the optical 
depth, which we define at the wavelength of $10~\mu$m  ($\tau_{10}$). A side result of 
this computation is the grain size distribution of the various dust species formed.

The chemo-dynamical description of the stellar wind, performed for all the selected points, is based on the schematization proposed by \citet{fg02, fg06}, 
in which the wind is assumed to expand isotropically from the surface of the star, 
until it reaches the dust condensation region, typically at temperatures below 
$\sim 1500$ K, where the physical conditions are such that the dust grains can 
form and grow. The relevant equations describing the dynamics of the wind and the
growth rate of the size of the various dust species, are extensively described and
discussed in \citet{ventura12}. The optical constants for the different dust 
species used in the present investigation are: amorphous carbon - \citet{zubko};
silicon carbide (SiC) - \citet{pegourie88}; solid iron - \citet{ordal}; silicates -
\citet{ossenkopf}; alumina dust - \citet{koike}.

These computations lead to the determination of the size reached by the
grains of the various dust species, the asymptotic velocity of the
wind, and the fraction of gaseous silicon, iron and carbon condensed into
dust \citep{ventura12}. These quantities, combined with the knowledge of the
surface chemical composition and of the mass loss rate of the star,
allow the computation of the production rates of the different
dust species and of the density stratification of the circumstellar envelope, according to the equations listed in section 5.2
of \citet{fg06}.

\begin{figure}
\begin{minipage}{0.48\textwidth}
\resizebox{1.\hsize}{!}{\includegraphics{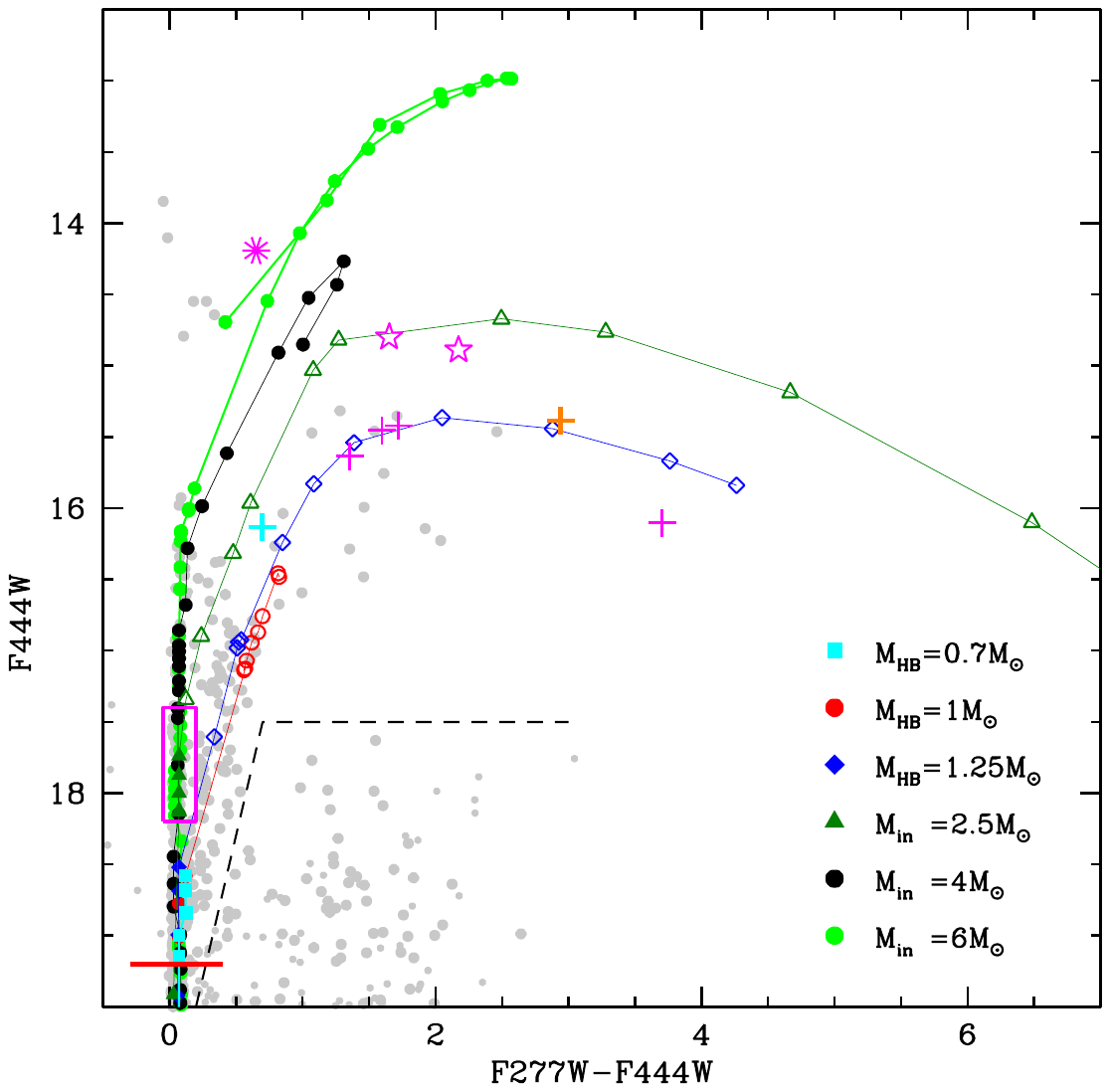}}
\end{minipage}
\vskip-60pt
\caption{Distribution of the evolved stellar population of Sextans A,
indicated with grey dots, in the $\rm (F277W-F444W, F444W)$ diagram. 
The thick, red, horizontal line indicates the location of the tip of 
the red giant branch predicted by the models. Sources 94328, 92104, 86434, 94477 investigated by \citet{boyer25} are indicated with magenta crosses; two stars exhibiting significant IR excess with star symbols; and a bright source, possibly a massive AGB undergoing hot bottom burning, as an asterisk. The crosses refer to the other two sources with MIRI-LRS spectroscopy by \citet{boyer25}: 90428 (orange), which shows evidence of SiC, and 90034 (cyan), which shows an oxygen-rich chemical composition. 
The colored points indicate various evolutionary phases of stars
of different initial mass. Full points refer to the phases during which the star 
is O-rich, whereas open symbols indicate C-rich phases. Note that for the 
$\rm 0.7~M_{\odot}$, $\rm 1~M_{\odot}$ and $\rm 1.25~M_{\odot}$ cases, which 
experience the helium flash, the masses refer to the start of the core helium 
burning phase (see text for details). The magenta box indicates the region partly 
populated by C-stars with little or no dust in the circumstellar envelope.
The region delimited by the dashed, black lines is dominated by background galaxies.
}
\label{fcmd1}
\end{figure}

For the selected points along the evolutionary sequences, we 
also calculated
the synthetic SED, using the radiative transfer code
DUSTY \citep{dusty}. In the first step the dust-free atmosphere for
each evolutionary stage considered is selected among the spectra
available in the COMARCS \citep{aringer09} and NEXTGEN \citep{nextgen}
libraries, for C- and M-stars, respectively. The final SED is
obtained by including the effects of dust, based on the results 
from stellar evolution + dust formation modelling. This enables the description of the SED temporal
evolution for stars with different progenitor masses during the AGB phase. 

Finally, to obtain the evolutionary tracks, the synthetic SEDs 
are convolved with the transmission curves of the NIRCam and MIRI filters,
which are used to derive the magnitudes in the various bands. We adopt a distance modulus $\mu_0=25.7$ mag \citep[see ][and references therein]{Yan2025} in order to compare the various evolutionary tracks with the observational data in different color-magnitude diagrams. With the $\rm E(B-V)=0.0374$ mag from \citet{Schlafly11}, and adopting a standard $\rm R_v$=3.1 reddening law, the models are reddened accordingly, and the extinction values in the JWST filters considered in this analysis are negligible.

\section{Understanding the NIRCam data}
\label{nircam}
As in other galaxies of the Local Group \citep{boyer11, boyer15a, boyer15b}, the AGB population of
Sextans A is made up of a majority of stars with little or no dust in
their surroundings, complemented by a minority of sources exhibiting a 
strong IR excess, indicating the ongoing formation of large
quantities of dust. The fraction of dusty objects is small, since dust formation occurs under relatively rare conditions: indeed the physical 
conditions required for significant dust production are reached only after 
the star experiences several thermal pulses. Moreover, the onset of dust formation enhances the mass-loss rate, thereby shortening the AGB lifetime \citep{flavia15a}.
In addition, metal-poor systems host relatively small stellar populations, which further reduces the probability of observing stars during these short-lived, dust-enshrouded evolutionary phases. As a consequence, the region of the parameter space associated with efficient dust formation is only sparsely populated in metal-poor galaxies such as Sextans A.

To study the dust-free objects, we use NIRCam data,
in particular the distribution of stars on the
$\rm (F277W-F444W, F444W)$ diagram, while the investigation of the
dusty sources is supported by the analysis of the MIRI fluxes. We adopted the $\rm (F277W-F444W, F444W)$ diagram because this filter combination provides a clear separation between dust-free and dust-enshrouded AGB stars, as it is highly sensitive to circumstellar dust emission at $\rm 4~\mu$m while still tracing the stellar photosphere at $\rm 2.8~\mu$m.
Other NIRCam color combinations provide a smaller dynamic range, thus a narrower color spread separating dust-free and dusty objects. These combinations suffer from stronger degeneracies between effective temperature and dust emission, making them less efficient for distinguishing the different AGB sequences.

The distribution of stars in Sextans A in the $\rm (F277W-F444W, F444W)$ 
diagram is shown in Fig.~\ref{fcmd1}, with grey points. Approximately $90\%$ 
of the sources populate an almost vertical sequence, at $\rm (F277W-F444W) \sim 0$ mag,
extending up to $\rm F444W \sim 16$ mag. To estimate the impact of Galactic foreground stars, we followed the method described by \citet{Blum06}, that uses the analysis of vertical sequences in near- and mid-infrared color-magnitude diagrams (such as $\rm (J-[3.6], [3.6])$), where foreground populations appear vertical due to the smearing of magnitudes over varying distances while maintaining constant colors, and found that approximately 12$\%$ of the stars above the tip of the RGB are Galactic foreground dwarfs and giants, with the relative contamination typically decreasing at fainter magnitudes. To estimate the Galactic foreground contamination we applied the same method by using the F090W and F150W fluxes, and we found that the contribution from foreground stars is below 5$\%$ of the sources in this sequence.
Consequently we conclude that the nearly vertical sequence is almost entirely 
populated by AGB stars belonging to Sextans A. In regard to background galaxies, 
we follow the discussion in \citet{boyer25} to locate them in the faint ($\rm F444W > 17.5$ mag) and red (0 mag  $\rm <F277W - F444W <$ 2 mag)  region of the $\rm (F277W-F444W, F444W)$ diagram, enclosed within the black, dashed lines in Fig.~\ref{fcmd1} \citep{Blum06}.

\begin{figure*}
\centering
\includegraphics[width=0.50\textwidth]{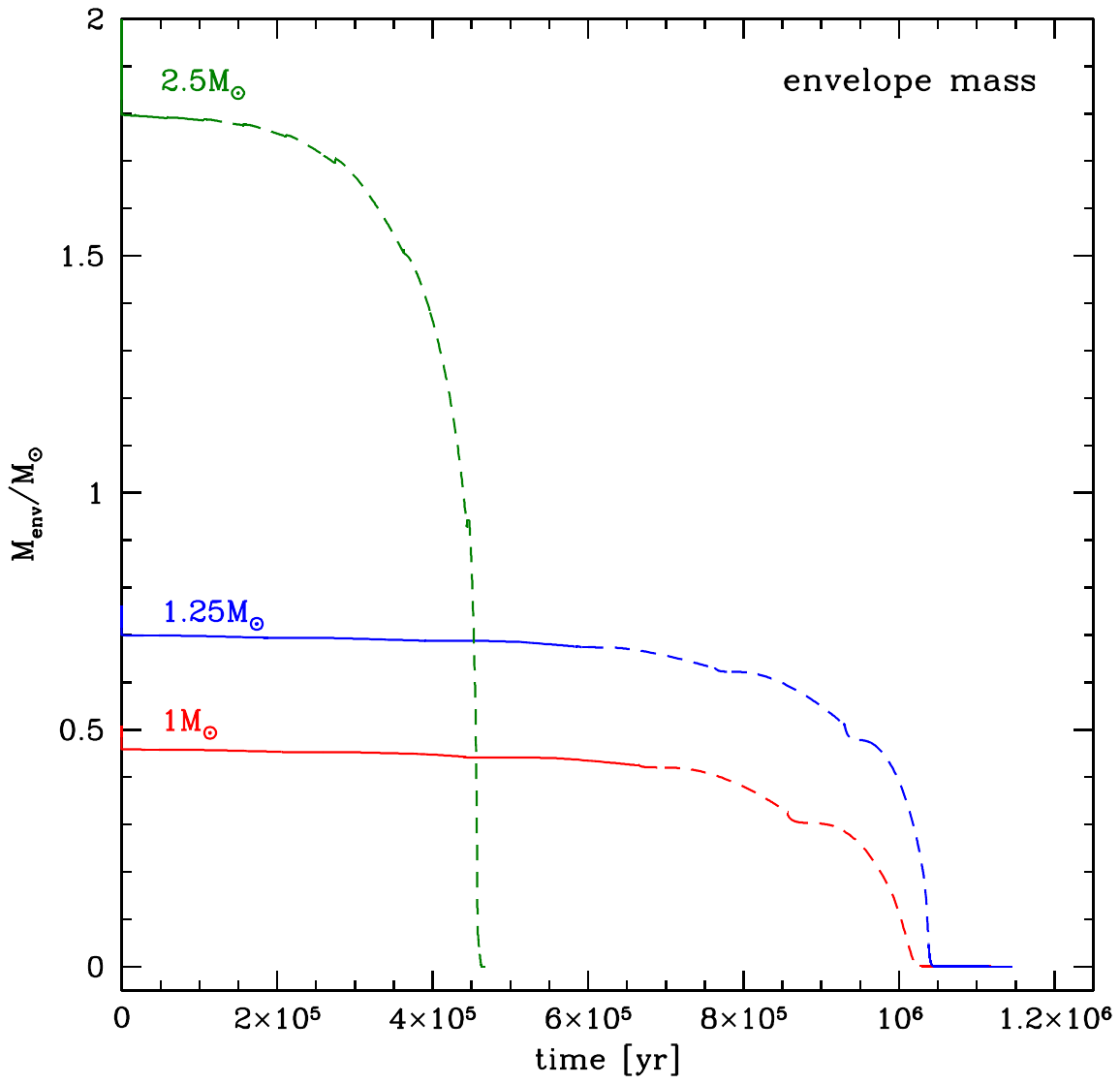}\hfil
\includegraphics[width=0.50\textwidth]{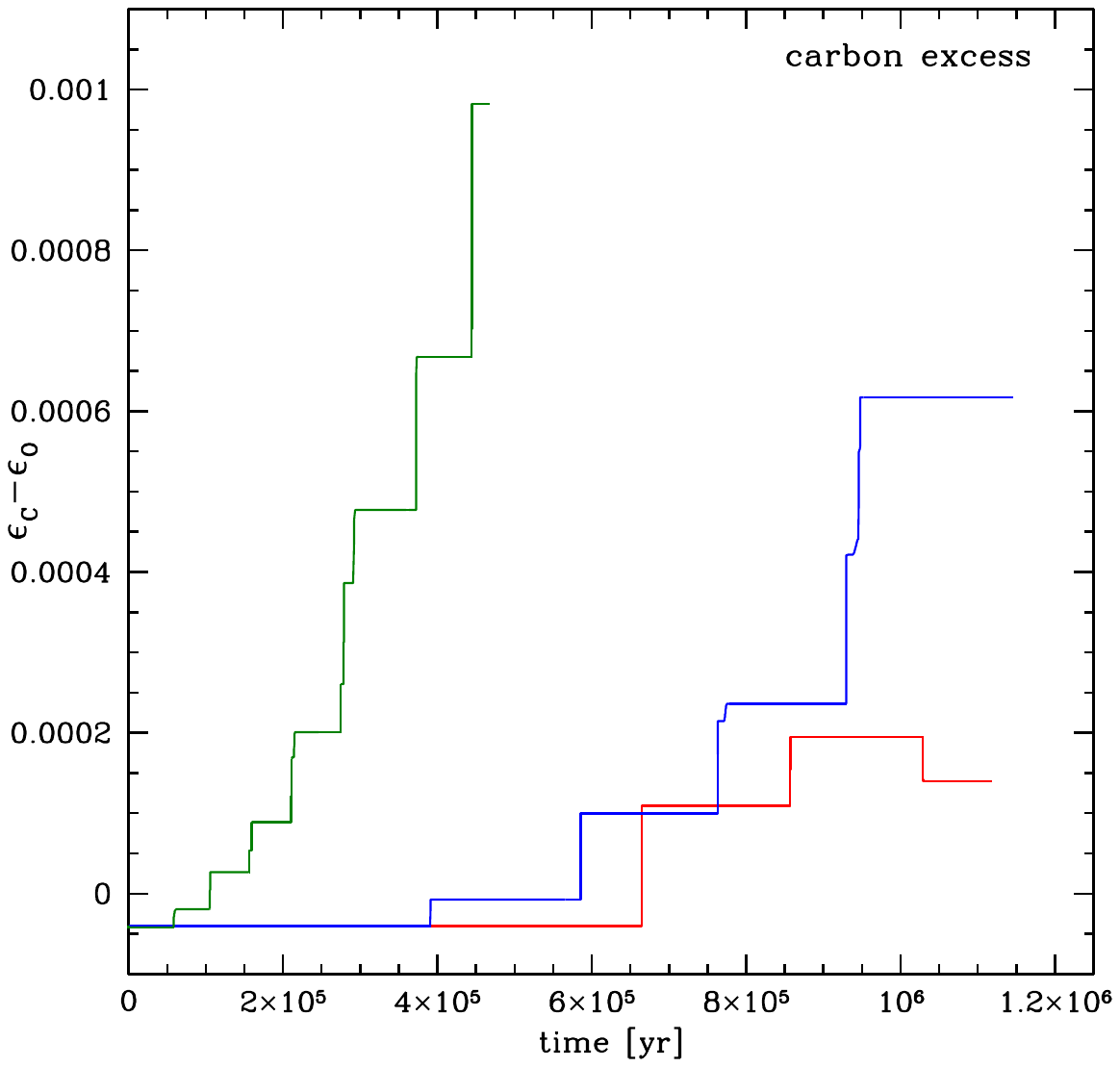}
\vskip-100pt
\includegraphics[width=0.50\textwidth]{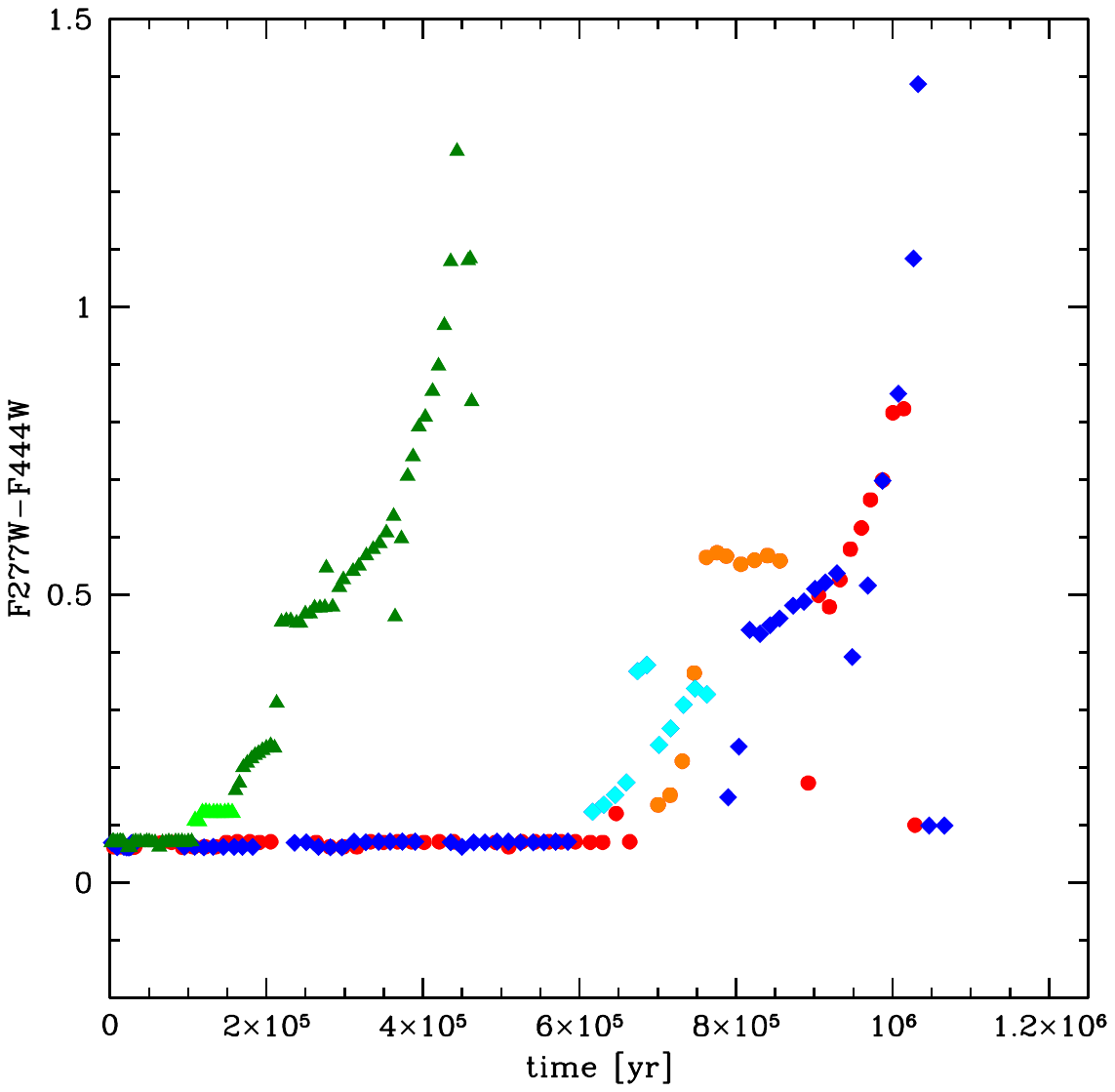}\hfil
\includegraphics[width=0.50\textwidth]{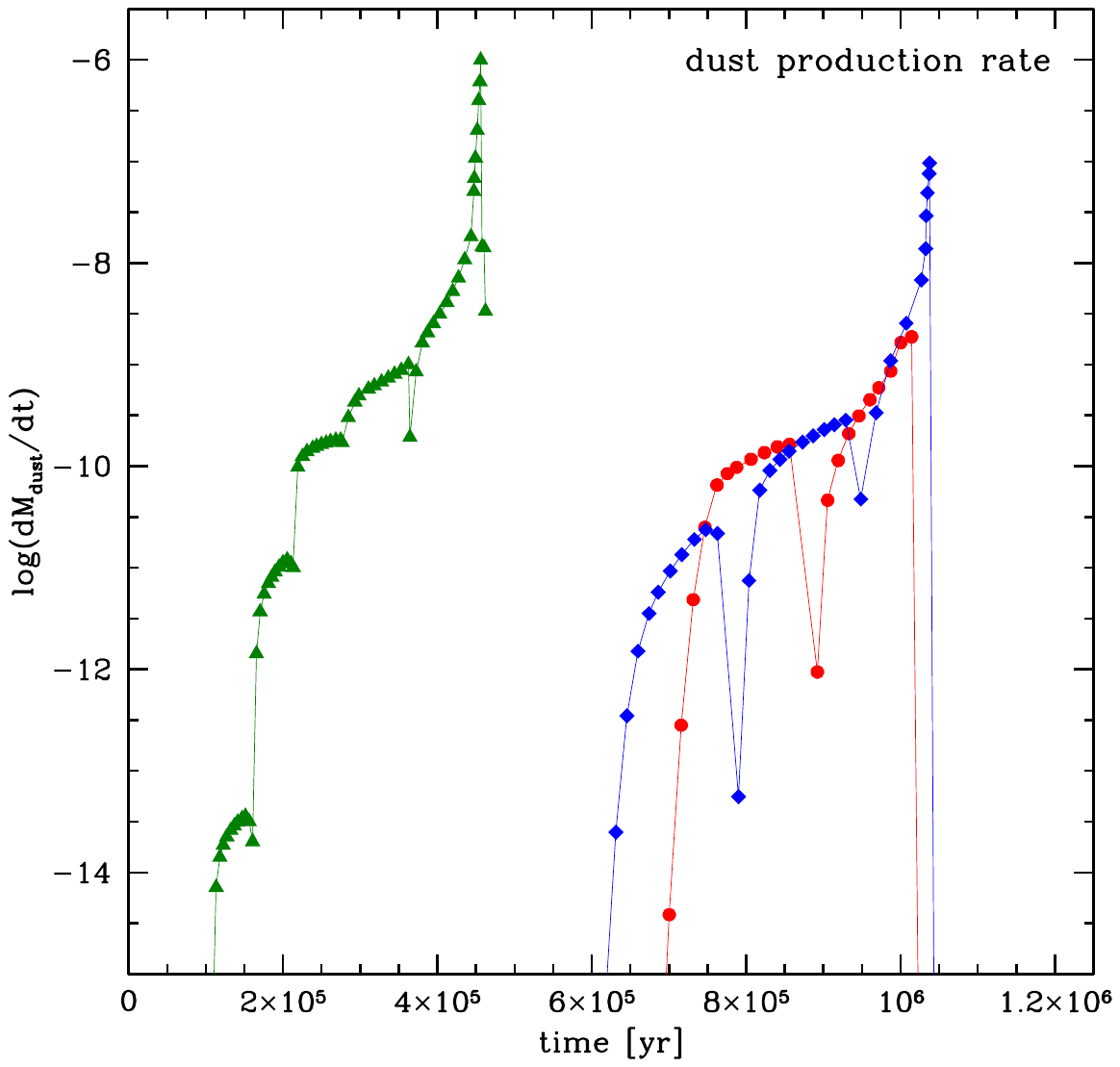}
\vskip-70pt
\caption{Variation of the mass of the envelope (top left panel),
carbon excess with respect to oxygen (top right), 
$\rm (F277W-F444W)$ color (bottom left) and carbon DPR (bottom right) 
during the AGB phase of model stars of initial mass $\rm 1~M_{\odot}$, 
$\rm 1.25~M_{\odot}$ and $\rm 2.5~M_{\odot}$. Times are counted from the ignition of the first thermal pulse, which
is experienced after 6.3 Gyr, 3 Gyr, half a Gyr since the formation
time for the three model stars considered. The solid and dotted part of the tracks in the top left panel indicate 
the O-rich and the C-rich phases, respectively. Orange full circles, cyan diamonds and light green triangles
in the bottom left panel refer, respectively, to selected evolutionary stages of 
the $\rm 1~M_{\odot}$, $\rm 1.25~M_{\odot}$, $\rm 2.5~M_{\odot}$ model stars, during the interpulse period 
following the achievement of the C-star stage. The vertical axis of the bottom 
left panel is limited to $\rm (F277W-F444W)$ = 1.5 mag, in order to preserve detail in the track morphology; a wider range would make the first part of the evolution indistinguishable.
}
\label{fevol}
\end{figure*}

In the following part of this section we first provide a general 
description of the expected excursion of the evolutionary tracks of 
stars of different mass across the aforementioned color-magnitude 
diagram. We then focus on the dust-free objects, which populate the 
blue side of the diagram, and discuss the use of the $\rm F444W$ flux
as luminosity indicator and the expected spectral type of these objects.
Finally, we describe the sources surrounded by dust and propose
a characterization of them in terms of mass and formation epoch of the
progenitor stars.

\subsection{The evolution of AGB stars along the $\rm (F277W-F444W, F444W)$ diagram}
\label{tracks}
Before entering the interpretation of the distribution of the Sextans A AGB stars
on the $\rm (F277W-F444W, F444W)$ diagram, it is important to note that
the modelling of the stars of mass below $\rm \sim 2~M_{\odot}$, which
experience the helium flash, requires an assumption regarding the
mass lost during the red giant branch (RGB) evolution, which determines the mass with
which the stars start the core helium-burning phase. For these stars,
the masses quoted in the text and in the figure captions 
refer to the start of the core helium-burning phase and will be 
labelled as $\rm M_{HB}$. The analysis of the morphology
of the HB in Globular Clusters, together with constraints from asteroseismology, shows that during the ascent of 
the RGB low-mass stars lose between $\rm 0.1~M_{\odot}$ and
$\rm 0.3~M_{\odot}$ of their envelopes \citep{miglio12, mcd15, tailo21, bro24}. Asteroseismic measurements provide the strongest quantitative evidence of RGB mass-loss and suggest that its efficiency increases with decreasing metallicity \citep{bro24}. As a consequence, $\rm M_{HB}$
is definitely smaller than the mass with which the stars formed, 
which will be labelled as $\rm M_{ZAMS}$. Accounting for the difference 
between $\rm M_{HB}$ and $\rm M_{ZAMS}$ is particularly relevant for 
the analysis of the evolution of stars of mass below $\rm \sim 1~M_{\odot}$, 
because the description of their AGB phase is highly sensitive to the mass of 
the envelope at the beginning of the AGB lifetime.

To understand the distribution of the stars in Fig.~\ref{fcmd1}, 
we follow the analysis of the AGB evolution discussed in 
\citet{ventura22}, who introduced three typical cases, namely: a) 
low-mass stars, which never reach the C-star stage, as they lose 
the envelope before the surface $\rm C/O$ exceeds unity; 
b) intermediate-mass stars, which become carbon stars (surface $\rm C/O$ ratio $>$ 1) after 
experiencing a series of thermal pulses and third dredge-up episodes \citep{iben}; c) massive 
AGB stars, which undergo the ignition of the hot bottom burning 
process \citep{booth93}, that prevents them from becoming carbon stars. 
The mass 
thresholds separating the three groups are sensitive to 
the metallicity of the stars. For the chemical composition considered 
in the present investigation, the minimum mass of the 
star at the start of the core helium-burning phase required to 
become a carbon star (thus separating type (a) from type (b) behavior) 
is $\rm M_{HB} \sim 0.8~M_{\odot}$ \citep{devika23}, while the minimum 
mass to ignite hot bottom burning at the base of the envelope is 
$\rm M_{ZAMS} \sim 3~M_{\odot}$ \citep{ventura13}.

To understand how the stars of different mass move across the
$\rm (F277W-F444W, F444W)$ diagram, we show in Fig.~\ref{fcmd1}, 
overlaid on the data points,
the evolutionary tracks of some selected model stars, to represent
the three cases discussed above. 
As an example of case a) we show 
the $\rm M_{HB}=0.7~M_{\odot}$ model star. As for the evolution 
of type b), regarding  the stars that reach the C-star stage, we show three cases, corresponding
to the masses $\rm M_{HB}=1~M_{\odot}$, $\rm M_{HB}=1.25~M_{\odot}$ and 
$\rm M_{ZAMS}=\rm 2.5~M_{\odot}$, to represent stars that accumulate different 
quantities of carbon in the surface regions, thus heterogeneous in 
the rate of carbon dust production experienced. Finally, the case c) is 
represented by the $\rm M_{ZAMS}=4~M_{\odot}$ and $\rm M_{ZAMS}=6~M_{\odot}$ model stars.

Regardless of mass, the evolutionary tracks draw approximately 
vertical sequences during the first part of the AGB evolution, as a 
consequence of
the gradual growth of the core mass, which reflects into
a rise in the luminosity \citep{blo93}, and thus in the $\rm F444W$ flux.
When dust formation in the wind begins, the SED is gradually
shifted toward the IR part of the spectrum and the evolutionary
tracks move to the red \citep{ester20, ester21}. This behavior 
is not shared by the $\rm 0.7~M_{\odot}$ model star, as the
stars belonging to group a), which evolve as oxygen-rich
objects during the entire AGB phase, produce dust at very
low rates. In the study of the dusty stars in the LMC, aimed
at interpreting the data taken with IRS onboard Spitzer,
\citet{ester20} concluded that some dust production, though
in limited quantities, occurs during the evolution of low-mass
stars in the O-rich phase. However, that result does not
hold in the present context, as the metallicities typical of
the stars in Sextans A are significantly lower than those
adopted by \citet{ester20}, as the scarcity of silicon
prevents the formation of significant quantities of dust. 
This conclusion is supported by studies of low-mass, low-metallicity 
stars in Galactic Globular Clusters \citep{boyer08,boyer09,mcd11a,mcd11b}, 
which similarly show very low dust production in such environments.

Regarding the model stars undergoing the evolution of
type b) discussed above, we distinguish between
the first part of the AGB evolution, during which they evolve
as M-type objects and the evolutionary tracks develop
vertically, and the phases following the
achievement of the C-star stage, when significant quantities
of carbon dust are formed and the evolutionary tracks move to the red. 
After reaching the C-star stage, the $\rm1~M_{\odot}$ model 
star evolves to the red side of the diagram until 
$\rm (F277W-F444W) \sim 1$ mag, and to brighter $\rm F444W$ 
fluxes up to $\rm F444W \sim 16.5$ mag; this is again related 
to the gradual luminosity increase, which grows from 
$\rm \sim 4800~L_{\odot}$ to $\rm \sim 5200~L_{\odot}$ during
the C-star phase.
The $\rm 1.25~M_{\odot}$ and $\rm 2.5~M_{\odot}$ model stars 
follow a different behavior. First, the horizontal extension
of the tracks is much wider and increases with the initial mass
of the star: the reddest points reached are 
$\rm (F277W-F444W) \sim 1, 4.5, 9.5$ mag for the 
$\rm 1~M_{\odot}, 1.25~M_{\odot}, 2.5~M_{\odot}$ cases, 
respectively. Furthermore, unlike the lower mass counterpart,
for the $\rm 1.25~M_{\odot}$ and $\rm 2.5~M_{\odot}$ cases 
the $\rm F444W$ flux first rises, then decreases after reaching 
a maximum value, which increases with the initial mass of the star.
Colors $\rm (F277W-F444W) > 5$ mag are reached
only by stars with mass in the $\rm 2-3~M_{\odot}$ range, and
during the very final evolutionary phases, when the
evolutionary time-scales become progressively shorter. The 
low probability of detecting stars evolving across this fast evolutionary
phase and the small AGB population of Sextans A
are the main reasons for the lack of sources in the 
$\rm (F277W-F444W) > 5$ mag region of the diagram.

These differences arise because the $\rm 2-3~M_{\odot}$ model
stars experience more third dredge-up events than the 
$\rm 1~M_{\odot}$ counterpart. As a result, larger quantities of carbon accumulate in the surface regions, triggering more intense dust production. Consequently, the peak of the
SED is shifted to wavelengths longer than $\rm 4.44~\mu$m
and the $\rm F444W$ flux decreases. This
behavior is shown in Fig.~\ref{fevol}, particularly in
the top right and bottom right panels, which show, respectively, the time
variation of the carbon excess with respect to oxygen 
(defined as $\rm \epsilon_C-\epsilon_O=(n_C-n_O)/n_H$) and of the
DPR of the three models discussed, derived following the method
described in section \ref{input}. 
We see that $\rm (\epsilon_C-\epsilon_O)$ grows until
$2\times 10^{-4}$, $6\times 10^{-4}$ and $10^{-3}$ for the
$\rm 1~M_{\odot}$, $\rm 1.25~M_{\odot}$ and $\rm 2.5~M_{\odot}$ 
cases, respectively, and the largest DPRs
are $2\times 10^{-9}$, $10^{-7}$ and $\rm 10^{-6}~M_{\odot}/$yr. 

The excursion of the evolutionary tracks to the red side
of the diagram, in the $\rm (F277W-F444W) > 1$ mag region, 
occurs during the final AGB phases, mostly the
very last inter-pulse, when the stars expand, the mass-loss rate increases significantly, and so does the DPR. When reaching this stage, the core mass of the stars has reached its final value, with which they enter the post-AGB and planetary nebula phases. The final core mass is higher for stars with larger initial masses. Given the tight connection
between core mass and luminosity, this implies that the
luminosity of the stars populating the 
1 mag $< \rm (F277W-F444W) < 4$ mag region of the diagram
grows with the initial mass of the star, ranging
from $\rm \sim 6000~L_{\odot}$, for the stars of mass slightly
above $\rm 1~M_{\odot}$, to $\rm \sim 18000~L_{\odot}$,
for $\rm 3~M_{\odot}$ stars. The vertical spread in 
the distribution of the C-rich sources on the red side
of the $\rm (F277W-F444W, F444W)$ diagram is therefore
due to the differences in the initial mass of the stars
that reach the C-star stage, with the additional
uncertainty associated with variability effects,
discussed earlier in this section.

As far as the group c) stars is concerned, we see in 
Fig.~\ref{fcmd1} that the evolutionary tracks of the $\rm 4~M_{\odot}$
and $\rm 6~M_{\odot}$ model stars first develop vertically, up to 
$\rm F444W \sim 16.5$ mag and $\sim 16$ mag, respectively,
then turn to the
red, when the activation of strong hot bottom burning leads to the
production of silicates at high rates. The tracks are
seen to return to the blue during the very final AGB
phases, because the hot bottom burning loses efficiency, the luminosity decreases, the mass-loss rate and the wind density decrease as well, thus reducing the DPR \citep{ventura12}.

\subsection{Dust-free stars}
We now focus on the objects populating the vertical strip
at $\rm (F277W-F444W) \sim 0$ mag, where the stars spend the majority 
(or the totality, as in the example of the $\rm 0.7~M_{\odot}$ star discussed
above) of their AGB lifetime. Since these stars are almost (or completely) 
dust-free, we refer to this region as the dust-free sequence (DFS).
In general, stars
of higher mass evolve on more massive cores, thus they
reach higher luminosities during the AGB phase. Therefore, as we move towards the brighter part 
of the DFS, we expect to find the progeny
of stars of higher mass, thus formed in more recent epochs.

The $\rm F444W $ flux is a reliable luminosity indicator for the
stars populating the DFS, which are characterised by similar
effective temperatures ($\rm T_{eff} \sim 3800-3900$ K). The comparison with 
the evolutionary tracks shows that the following relationship
between the luminosity of the star and the $\rm F444W$ flux holds:

\begin{equation}
\log \left( \frac{L}{L_{\odot}} \right)
= -0.35\, F444W + 10.15 .
\end{equation}
The relationship given by Eq. 1 is a linear fit to 
the synthetic photometry and bolometric luminosities derived 
from the evolutionary tracks. It is important to note that this 
diagnostic is valid only for stars in the DFS, for which
the peak in the SED is located at wavelengths significantly
shorter than $4~\mu$m. This same calibration could hardly be
applied to the sources populating the $\rm (F277W-F444W) > 1$ mag region, 
since dust reprocessing causes the F444W flux to deviate 
from the bolometric trend given by Eq. 1.

With regard to the chemical spectral type, the models indicate
that most of the sources populating the DFS are
M-type objects. This is definitely true for the lowest masses,
those following the behavior of type a) mentioned above, 
and for the massive AGBs experiencing hot bottom burning (type c)), which never
become carbon stars. The stars that reach the C-star
stage move to the red shortly after the surface 
$\rm C/O$ exceeds unity. Here a distinction is required
according to the initial mass of the star, because the
residual mass of the envelope when the C-star phase is reached
changes with the initial mass of the stars: as shown
in the top left panel of Fig.~\ref{fevol}, the residual envelope
mass when the C-star phase begins decreases from
$\rm \sim 1.75~M_{\odot}$ for the $\rm M_{ZAMS}=2.5~M_{\odot}$
star to $\rm \sim 0.45~M_{\odot}$ in the $\rm M_{HB}=1~M_{\odot}$
case.

In stars descending from $\rm M_{HB}<1.5~M_{\odot}$ progenitors,
dilution of the material dredged-up 
during the third dredge-up with the thin (in mass) convective envelope
favors a significant increase in the carbon excess with
respect to oxygen, which, in turn, leads to high rates of
dust production: these stars are expected to move to the red 
as they become carbon stars, thus they evolve within the
DFS during the M-type phase only. On the other
hand, the higher mass counterparts become C-stars after
experiencing a number of thermal pulses and third dredge-up episodes, when only
a small fraction of the envelope mass has been lost. In this
case, the dilution between the dredged-up material
processed by helium burning and the external convective
zone is highly efficient, so that, when the C-star is reached,
the carbon excess is small and little dust formation takes
place. Unlike the lower mass counterparts, these sources
continue to evolve along the DFS during the
early evolutionary phases after becoming C-stars, and evolve
to the red only in a later phase, after developing a
significant carbon-to-oxygen excess. The only exception 
to this behavior is found for the stars of initial mass close
to the minimum threshold required for hot bottom burning ignition, whose SED is shifted to the mid-IR from the onset of the C-star phase. This peculiar behavior is driven by their lower surface gravities than their lower mass counterparts: indeed stars in this mass range reach higher luminosities, which leads to a significant expansion of the structure. The resulting lower surface gravity makes the outer envelope less gravitationally bound, thereby facilitating an earlier onset of intense mass-loss. This creates the dense and cool circumstellar conditions necessary for dust grains to condense more efficiently, even at the very beginning of the C-star stage.

In summary, we conclude that the DFS
observed in Fig.~\ref{fcmd1} is mostly populated by oxygen-rich
stars, but it also harbors a smaller fraction of C-stars,
descending from progenitors with masses in the $\rm 1.5-2.5~M_{\odot}$
range. The results from stellar evolution modelling indicate
that the luminosities of the model stars in the mass interval 
given above, upon reaching the C-star stage, span the
$\rm 5000-10000~L_{\odot}$ range, which corresponds to 
17.4 mag $<$ $\rm F444W <$ 18.2 mag. This region of the diagram is
indicated with a magenta box in Fig.~\ref{fcmd1}. C-stars are not 
expected at fainter $\rm F444W$ fluxes nor in the 
brighter part of the DFS, in the region above 
$\rm F444W \sim 17.4$ mag.

On the basis of the present analysis, we
identify three distinct regions in the vertical DFS
of the $\rm (F277W-F444W, F444W)$ diagram. In the
$\rm F444W > 18.2$ mag part we find a mixture of 
O-rich stars, different in mass and formation epoch,
evolving across the AGB phase. Most of these sources,
for reasons connected to the shape of the initial mass 
function and
to the evolutionary time scales of this phase, descend
from stars of mass $\rm M_{HB} < 1.5~M_{\odot}$, formed
earlier than $\sim 1.7$ Gyr ago. The brighter region of the strip, with 
17.4 mag $<$ $\rm F444W <$ 18.2 mag, harbors both M-type
objects and C-stars descending from 
$\rm 1.5~M_{\odot}-2.5~M_{\odot}$ progenitors, which formed
between half a Gyr and 1.5 Gyr ago. Based on the 
relative duration of the C-rich and O-rich phases, and taking 
into account the initial mass function and the AGB time-scales of stars of
different mass, we estimate that if the star formation
rate did not change significantly during those epochs, the fraction of
C-rich stars in this region of the diagram is $\sim 20\%$.
Finally, in the bright region of the diagram, at $\rm F444W < 17.4$ mag,
we expect to find only stars of mass above $\rm 3~M_{\odot}$,
during the evolutionary phases following the ignition of hot bottom burning.

It is important to note that our sample is based on single-epoch observations. As shown, for instance, by the DUSTiNGS survey with Spitzer \citep{boyer15b}, the magnitudes of AGB stars at wavelengths $\lambda \sim 3$–$4~\mu$m can vary with amplitudes up to 0.5 mag. This introduces additional
uncertainty in the interpretation of the data reported in Fig.~\ref{fcmd1}. 
In particular, while we consider it most likely that C-stars along the
DFS populate the $\rm 17.4 < F444W < 18.2$ mag strip, the possibility
that a few carbon-rich objects are located within $\sim 0.2$ mag outside the F444W thresholds given above cannot be ruled out.
However, because the AGB lifetime is much longer than the pulsation period, these effects average out across the total population, and the relative proportions of stars in each region (e.g., the $\sim$ 20$\%$ C-star fraction estimate) remain statistically valid for characterizing the intermediate-age population. The single-epoch 
nature of the observations does not affect the validity of Eq. 1,
as far as sources populating the DFS are considered. Assuming a typical variability amplitude of $\pm$ 0.25 mag for these sources, the uncertainty propagated into the calculated luminosity is $\sim$0.09 dex. Despite this single-epoch scatter, the relation provides a robust estimate of the average luminosity of the DFS population, as the dispersion is smaller than the typical range of luminosities spanned by the AGB phase.

\subsection{Dusty objects}
\label{dusty}
E. Tarantino et al. (2026, in preparation) show that Sextans A harbors $\sim$ twenty very dusty AGB stars, which are visible in Fig.~\ref{fcmd1} as stars populating the region of the diagram redward of the DFS discussed above. As discussed
earlier in this section that this part of the color-magnitude
diagram harbors stars with carbonaceous dust in their surroundings,
whose formation is driven by the significant quantities of
carbon accumulated in the surface regions.

The majority of these objects, in particular those populating the
0 mag $< \rm (F277W-F444W) < 1$ mag region, are interpreted
as the progeny of stars of mass below $\rm \sim 1.5~M_{\odot}$,
which have recently become C-stars. This understanding is consistent
with the morphology of the evolutionary tracks in Fig.~\ref{fcmd1},
which in fact overlap with the positions of these sources.

The redder objects found in the region 1 mag $< \rm (F277W-F444W) < 4$ mag
exhibit a significantly higher IR excess, consistent with their
identification as stars that have experienced a sequence
of third dredge-up episodes that favored the rise in the surface
$\rm C/O$ up to $\sim 10$. Based on the comparison between
the position of these sources and the evolutionary tracks reported
in Fig.~\ref{fcmd1}, we deduce a dominant presence of stars descending from 
$\rm M_{HB} < 1.5~M_{\odot}$ progenitors, plus the two brightest ones,
indicated with magenta stars in the figure, at $\rm F444W \sim 15$ mag, 
which likely descend from progenitor stars with $\rm M_{ZAMS} \sim 2.5~M_{\odot}$, formed $\sim$ 0.5 Gyr ago.

In Fig.~\ref{fcmd1} we indicated with magenta and orange crosses 
five out of the six sources with available MIRI-LRS 
spectroscopy, recently studied by \citet{boyer25}, which populate
the $\rm (F277W-F444W) > 1$ mag region of the CMD. The position
of these objects is well reproduced by the track of the 
$\rm 1.25~M_{\odot}$ model star; however, taking into account
possible variability effects, we can reasonably assume progenitor masses 
for these stars in the $\rm 1.15~M_{\odot} < M < 1.5~M_{\odot}$
range.
The five objects, which 
according to our interpretation represent the final evolutionary phases of stars
formed $\sim 2-3$ Gyr ago, ideally trace a sequence of stars
enriched in carbon, which are currently producing carbonaceous dust 
at different rates, from $\rm 10^{-8}~M_{\odot}/$yr to 
$\rm 10^{-7}~M_{\odot}/$yr. Most of the dust is released in the
form of amorphous carbon, while the contribution of SiC is not expected
to exceed $\sim 5\%$: this is consistent with the results from
\citet{boyer25}, who find clear evidence of SiC only in the SED
of source 90248, which is indicated in orange in Fig.~\ref{fcmd1}. 
The JWST/LRS spectra of these five sources clearly show the presence of 
the acetylene (C$_2$H$_2$) absorption band at 7.5 $\mu$m, confirming their carbon star nature. 
Their strength increases with the star color, indicating a larger 
carbon excess as expected from the $\rm 1.25~M_{\odot}$ model (see Fig. \ref{fevol}). 

This interpretation does not apply for source 90034, indicated with a cyan cross in Fig.~\ref{fcmd1}, which exhibits a significantly lower IR excess. The strong water feature around 6.5 $\mu$m and the lack of 
C$_2$H$_2$ in the JWST/LRS spectrum of this object indicates that it is not
a carbon star. Furthermore, the presence of a featureless dust continuum was 
associated by \citet{boyer25} to the presence of metallic iron dust.
A dominant contribution from iron dust with respect to silicates
can be obtained during the very final evolutionary phases of metal-poor,
massive AGB stars, because the hot bottom burning experienced is so efficient to cause a severe 
depletion of surface oxygen and magnesium \citep{flavia18}, 
which, in turn, leads to a dramatic decrease in the rate at which 
silicates are formed \citep{marini19}; on the 
other hand, the formation of solid iron is not affected by the
lack of gaseous silicon and magnesium. Therefore, we tentatively identify
90034 as the descendant of a $\rm \sim 4-5~M_{\odot}$ star, in agreement
with the large luminosity of $\rm \sim 20000~L_{\odot}$ estimated by 
\citet{boyer25} for this object.

In the brightest part of the diagram we note a paucity of objects with 14 mag $< \rm F444W < 15$ mag and $\rm (F277W-F444W) \sim 0$ mag, which 
may be identified as either red supergiant stars (RSGs) or massive AGB stars.
The AGB interpretation is the more likely for the source marked with a magenta asterisk. Its position on the diagram, which is reproduced by the evolutionary track of the $\rm 6~M_{\odot}$ 
model star, deviates significantly from the DFS, suggesting that 
either it is surrounded by silicates and alumina dust, or,
similarly to source 90034, by metallic iron dust. The latter hypothesis 
is consistent with recent JWST/LRS results, which indicate that silicate features may be rare among the AGB sources of this galaxy \citep{boyer25}. We note, however, that this star is near the saturation limit of the NIRCam detector, and possible non-linear effects may impact its measured magnitude and color. Despite this caveat, it remains a valid candidate for being a massive AGB star.

While this classification provides 
a characterization of the sources on the red side of
the diagram in terms of the progenitor mass and the
dust produced in their circumstellar envelope, 
a more detailed analysis of the dust mineralogy surrounding
these objects requires the MIRI data,
as the silicates and SiC features
are found in the $10-20~\mu$m spectral region.
This analysis will be addressed in the next section.

\begin{figure*}
\centering
\includegraphics[width=0.50\textwidth]{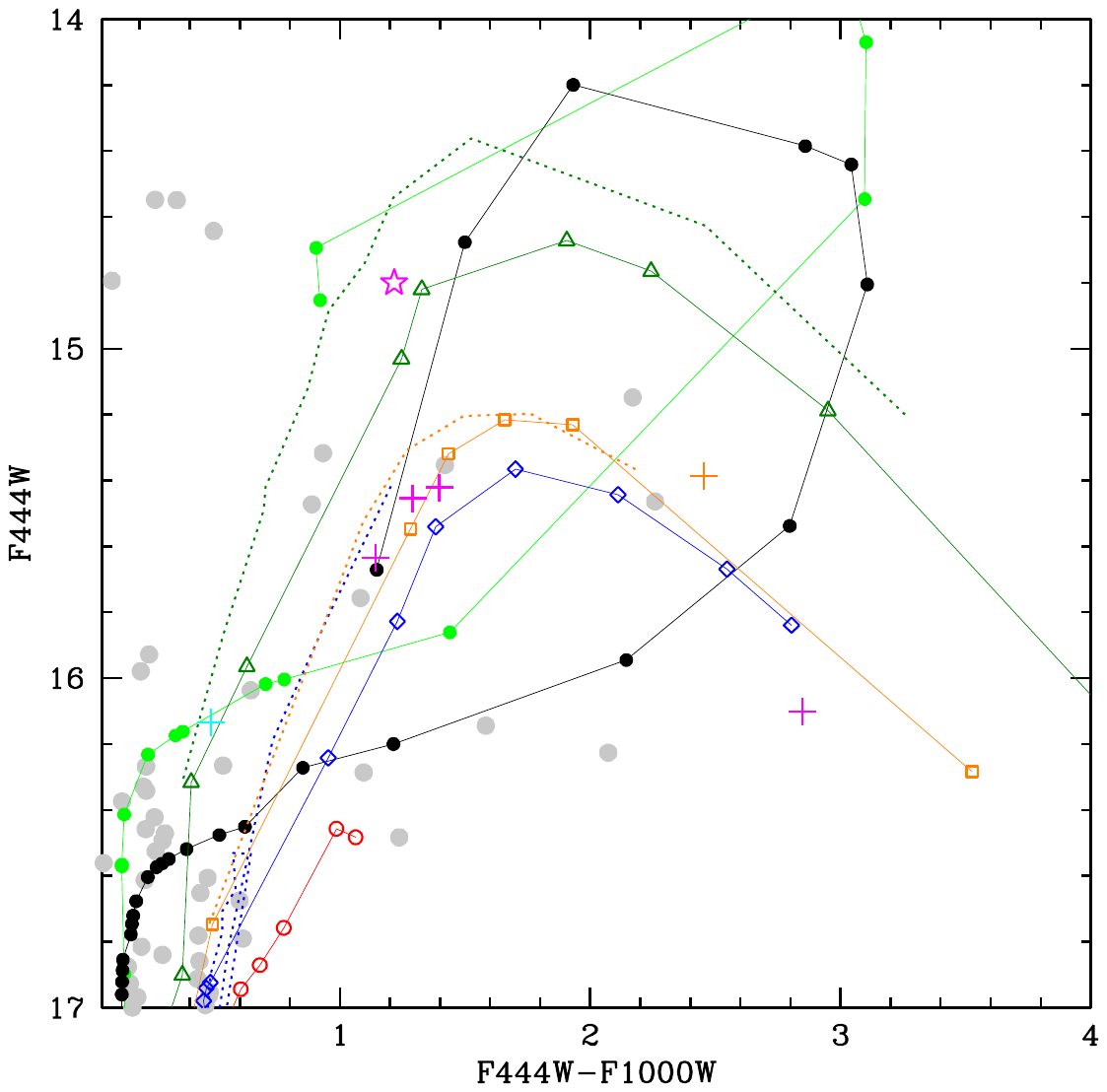}\hfil
\includegraphics[width=0.50\textwidth]{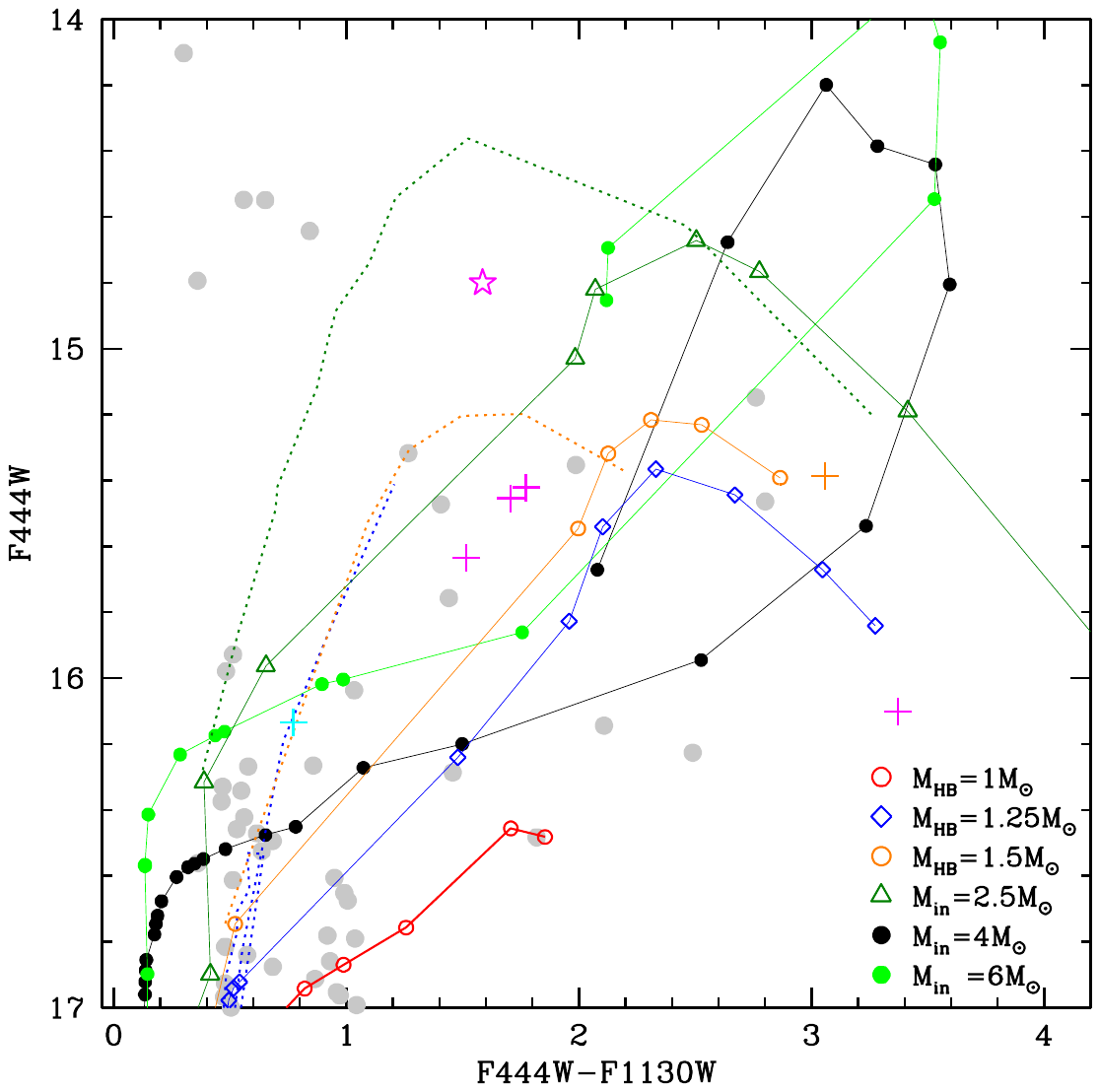}
\vskip-70pt
\caption{Distribution of the evolved stellar population of Sextans A,
indicated with grey dots, in the $\rm (F444W-F1000W, F444W)$ (left panel) and
$\rm (F444W-F1130W, F444W)$ (right panel) diagrams. Crosses, stars, asterisks and model symbols connected by solid lines, have the same meaning as in Fig.~\ref{fcmd1}. Dotted lines indicate the evolutionary of metallicity $\rm Z=4\times 10^{-4}$.}
\label{fmiri}
\end{figure*}

\section{Interpretation of the MIRI imaging data}
\label{miri}
For some of the sources shown in Fig.~\ref{fcmd1} MIRI photometry in the
F770W, F1000W, F1130W and F1280W bands is available. To further validate the
general understanding presented in the previous section, particularly
in the interpretation of the reddest objects discussed in section
\ref{dusty}, we analyze the distribution of the stars in Sextans A
in the $\rm (F444W-F1000W, F444W)$ and $\rm (F444W-F1130W, F444W)$
diagrams. We consider these two diagrams only, because the COMARCS
atmospheres \citep{aringer09} of carbon stars used as input for 
the code DUSTY do not account 
for the depressions of the flux in the $7-8~\mu$m and $12-13~\mu$m spectral
regions related to the presence of $\rm C_2H_2$ molecules \citep{sloan16}:
the synthetic approach would clearly overestimate the F770W and F1280W fluxes,
with the exception of the stars surrounded by large quantities 
of dust, where the radiation released from the
photosphere is entirely reprocessed by the dusty region of the
circumstellar envelope.

The distribution of the AGB sources of Sextans A on the aforementioned color-magnitude diagrams is shown in Fig.~\ref{fmiri}. 
The evolutionary tracks of selected model stars are overplotted on the data points and are particularly useful for the present analysis.

On the blue, faint side of the diagrams we note the presence of
$\sim$ twenty objects at $\rm (F444W-F1000W) \sim 0.1$ mag and 16.2 mag $< \rm F444W < 17$ mag,
which were found to populate the upper part of the DFS
shown in Fig.~\ref{fcmd1}. The position of these objects is
well reproduced by the evolutionary tracks of $\rm 4-7~M_{\odot}$
stars, which follow an approximately vertical path until the
strongest hot bottom burning conditions are reached and formation of silicates
takes place. We therefore interpret these objects as the progeny
of massive AGBs, formed between 50 and 200 Myr ago, currently
undergoing negligible or soft hot bottom burning at the base of the envelope. 

The behavior of the tracks of the stars that reach the C-star stage,
in this case represented by the $\rm 1.25~M_{\odot}$, $\rm 1.5~M_{\odot}$
and $\rm 2.5~M_{\odot}$ cases, is qualitatively similar to that
discussed in Fig.~\ref{fcmd1}: a) the tracks develop redward, owing to the
gradual increase in the surface carbon, which enhances the DPR; b) towards 
the final evolutionary phases, the F444W flux, after reaching
a maximum value that depends on the progenitor mass, eventually decreases as the peak of the SED of very dusty stars is found at 
wavelengths above $5~\mu$m. 

Massive AGBs experiencing hot bottom burning, here represented by the $\rm 4~M_{\odot}$ 
and $\rm 6~M_{\odot}$ model stars, follow a different behavior.
The evolutionary tracks follow an anti-clockwise trajectory,
which can be explained on the basis of the arguments presented at the
end of section \ref{tracks}: the initial excursion to the red side
of the diagrams is driven by the progressively increasing
production of silicates characterizing the first part of the AGB evolution, 
while the return to the blue begins after the peak of the strength of hot bottom burning 
is reached and dust formation occurs at lower and lower rates.

The bright blue sources located in the 14 mag $< \rm F444W < 15$ mag and $\rm (F277W-F444W) \sim 0$ mag region (see Fig.~\ref{fcmd1}) were identified as either RSGs or massive AGB 
stars in section \ref{dusty}. However, their distribution in the MIRI color-magnitude diagrams (Fig.~\ref{fmiri}) can be used to assess their nature. Indeed the position of these sources lie notably blueward of the AGB evolutionary tracks, which favours their interpretation as RSGs, with the possible exception of the single source marked with an asterisk, which remains a strong massive AGB candidate.

As discussed in the previous section, source 90034 can be identified 
as the progeny of a massive AGB star, based on the large luminosity 
and the characteristics of the SED. Although the final points of the
evolutionary tracks of the $\rm 4~M_{\odot}$ model star shown
in Fig.~\ref{fmiri} correspond to luminosities $\sim 30\%$ higher and colors $\sim$ 1 mag redder 
than those estimated by \citet{boyer25} for this source,
it is still plausible that 90034 originates from
a $\rm \sim 4~M_{\odot}$ progenitor that is currently evolving through the very final AGB phases. These stages are not included in the evolutionary tracks shown in Fig.~\ref{fmiri} due to numerical convergence limitations.

The characterization of the individual sources based on the results
reported in the two panels of Fig.~\ref{fmiri} is slightly different 
from that deduced on the basis of the analysis of Fig.~\ref{fcmd1}. 
The results of Fig.~\ref{fcmd1} show that the position of the
sources in the 1 mag $< \rm (F277W-F444W) < 4$ mag region of the diagram is 
nicely reproduced by the evolutionary tracks of model stars of 
mass $\rm M_{HB}=1, 1.25~M_{\odot}$ and $\rm M_{ZAMS}=2.5~M_{\odot}$.
On the other hand, we note in Fig.~\ref{fmiri} that 
the position of the same sources in the $\rm (F444W-F1000W, F444W)$ and 
$\rm (F444W-F1130W, F444W)$ diagrams is not well reproduced by the 
evolutionary tracks of the same model stars. This is reflected in an attribution 
of slightly higher masses to the progenitors of the individual sources, 
with respect to those deduced in
the previous section. For instance, if we consider the five reddest objects
studied by \citet{boyer25}, indicated with magenta and orange crosses in Fig.~\ref{fcmd1} and
\ref{fmiri}, we concluded in the previous section that they were the 
progeny of $\rm M_{HB} = 1.2-1.3~M_{\odot}$ stars, whereas, based on 
their position on the $\rm (F444W-F1000W, F444W)$ and 
$\rm (F444W-F1130W, F444W)$ diagrams, we would derive a progenitor mass 
of $\rm M_{HB} \sim 1.5~M_{\odot}$ and $\rm M_{ZAMS} \sim 2~M_{\odot}$, 
respectively. A further example is provided by the
bright, red source indicated with a magenta star in Fig.~\ref{fmiri}, which
is associated with a progenitor mass of $\rm M_{ZAMS} = 2.5~M_{\odot}$ when considering the
$\rm (F277W-F444W, F444W)$ and $\rm (F444W-F1000W, F444W)$ diagrams, 
while the analysis of the $\rm (F444W-F1130W, F444W)$ diagram suggests
a progenitor mass around $\rm M_{ZAMS} = 3~M_{\odot}$.

To understand the reasons for the discrepancy between the conclusions
drawn from the analysis of the color-magnitude diagrams obtained with the NIRCam 
and MIRI filters, we note that both the F1000W and F1130W fluxes of 
carbon stars are strongly affected by the feature in the SED 
centered at $11.3~\mu$m, associated with the formation of SiC.
We first consider the possibility that the choice of the optical constants related to
SiC could play a role in this context. The evolutionary tracks shown in Fig.~\ref{fmiri},
which are more consistent with the interpretation of dusty objects given 
in section \ref{dusty}, are obtained using the \citet{pegourie88} optical constants. 
Use of the optical constants by \citet{laor93} leads to colors that are too red by 
$\sim 0.5$ mag with respect to the position of the data points, which cannot be 
reproduced by any of the evolutionary tracks of $\rm 1.25-3~M_{\odot}$ stars, which 
we have seen to reach the C-star stage.

We are left with the option that the present modelling, while well reproducing 
the gradual shift to longer wavelengths of the SED of stars as they evolve toward the 
final phases of the AGB evolution, overestimates the amount of SiC formed.
This is supported by the fact that among the five carbon stars
identified by \citet{boyer25}, only one shows clear evidence
of SiC in its JWST/LRS spectrum. Indeed, the presence of SiC in the
circumstellar envelope is irrelevant for the location of the stars
on the $\rm (F277W-F444W, F444W)$ diagram, while it affects the 
position on the $\rm (F444W-F1000W, F444W)$ and $\rm (F444W-F1130W, F444W)$
diagrams. We tested the possibility that lower values of the sticking coefficient 
$\rm \alpha_{Si}$ of gaseous silicon
on solid SiC particles could partly inhibit the formation of SiC, thus leading to
less prominent SiC features, hence redder $\rm (F444W-F1000W)$ and $\rm (F444W-F1130W)$
colors. However, we find that the results are largely independent of the choice of $\rm \alpha_{Si}$, because SiC easily reaches saturation, being the most stable compound formed in the circumstellar envelopes of carbon stars \citep{fg06}.

A possible way to decrease the efficiency of the formation of SiC is to assume 
that the old stellar component of Sextans A is composed of stars whose metallicity 
is lower than the $\rm [Fe/H]=-1.4$ value adopted here. This would be consistent
with the analysis by \citet{sakai96}, who derived a metallicity 
$\rm [Fe/H] \sim -1.8$, based on the morphology of the RGB of the galaxy, 
and was recently adopted by \citet{boyer25}, who assumed an initial metallicity of 
Z=0.0004 for their computations. This hypothesis would also be
consistent with the results by \citet{sloan12}, who compared the
height of the SiC features in carbon stars in the galaxies
Leo I, Carina ($\rm [Fe/H] \sim -1.4$) and Sculptor ($\rm [Fe/H] \sim -1.8$). 
While the SED of the C-stars in the first two galaxies exhibit a SiC feature
with a 0.2 excess with respect to the continuum, the counterparts in
Sculptor are characterised by a SiC-free SED.

A lower $\rm [Fe/H]$ would significantly affect the formation
of SiC, as the amount of the latter dust species that can be
formed is tightly connected to the silicon content available in
the surface regions of the stars \citep{fg06}.
To quantify the effects of a metallicity change, we calculated evolutionary
sequences of $\rm Z=4\times 10^{-4}$ model stars of initial mass in the 
$\rm 1.25-2.5~M_{\odot}$ range. More metal-poor stars of a given initial mass
evolve on more massive cores than their higher-metallicity counterparts, 
thus they reach brighter luminosities. The formation of carbon dust
is less efficient in metal-poor environments, owing to the hotter effective
temperatures, which partly inhibits the formation of dust in the circumstellar
envelope. 

On the $\rm (F277W-F444W, F444W)$ diagram shown in Fig.~\ref{fcmd1} the evolutionary tracks
of the $\rm Z=4\times 10^{-4}$ model stars would be slightly brighter than
those corresponding to the $\rm Z=10^{-3}$ case, and in general
they do not reach the extremely red colors attained by $\rm Z=10^{-3}$ stars.
The same interpretative approach adopted in the previous section would
allow to characterize three out of the five sources (sources 94328, 92104, 86434) identified as carbon stars and
discussed in \citet{boyer25}, 
with
progenitor's masses slightly smaller than those deduced in the
previous section. 

For what concerns the MIRI color-magnitude diagrams, the $\rm Z=4 \times 10^{-4}$ evolutionary tracks are significantly bluer than the $\rm Z=10^{-3}$ ones
(see the dotted lines in Fig.~\ref{fmiri}), owing to the lower
amount of SiC formed, which decreases the $\rm F1000W$ and $\rm F1130W$ fluxes.
The metal-poor tracks reported in Fig.~\ref{fmiri} lay on the blue side of the 
MIRI color-magnitude diagrams with respect
to the position of the sources populating the 1 mag $< \rm (F277W-F444W) < 4$ mag region in Fig.~\ref{fcmd1}, thus the progenitor masses
estimated for these objects would be smaller than those deduced in section \ref{dusty}.
We therefore find a reverse situation with respect to the results obtained
when applying the $\rm Z=10^{-3}$ tracks, which we showed to be too red to explain the same
sources. From this comparative analysis
we conclude that consistency between the results based on the NIRCam and MIRI data is obtained when an intermediate metallicity between $\rm Z=4\times 10^{-4}$ and $\rm Z=10^{-3}$ is adopted.
Overall, this interpretation provides further support to the hypothesis that the intermediate-age stellar population of Sextans A is characterized by metallicities slightly lower than the $\rm [Fe/H] \sim -1.4$ value adopted in our reference models.

\begin{figure}
\begin{minipage}{0.48\textwidth}
\resizebox{1.\hsize}{!}{\includegraphics{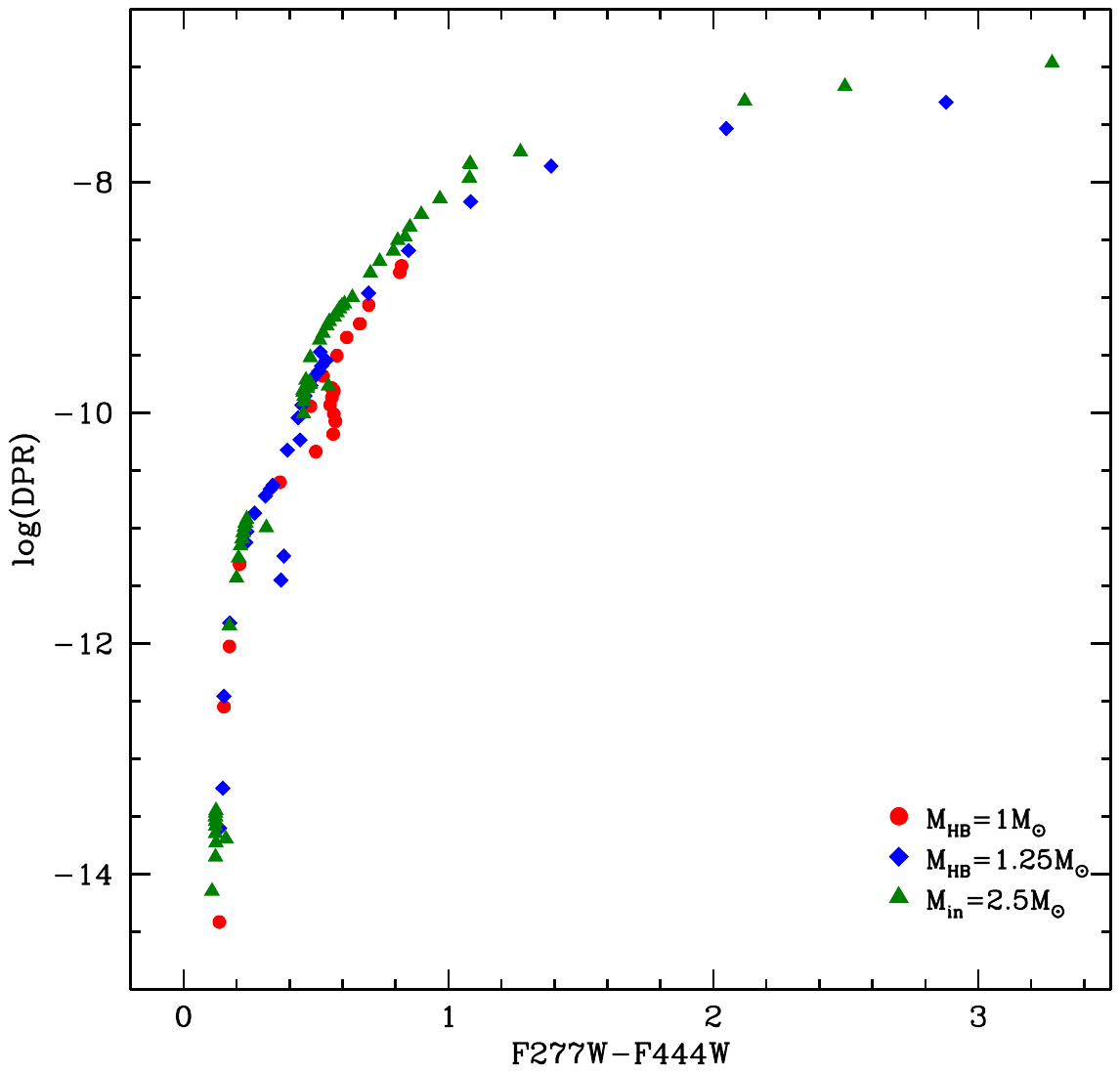}}
\end{minipage}
\vskip-60pt
\caption{Variation of the dust production rate as a function of the $\rm (F277W-F444W)$ color
during the AGB evolution of the same model stars 
shown in Fig.~\ref{fevol}.}
\label{fcoldpr}
\end{figure}

\section{The dust production rate}
\label{dpr}
While the mineralogy of the dust surrounding the stars
is primarily constrained by using MIRI data,
we base the discussion of the DPR on the NIRCam colors, because MIRI photometry is available
for only part of the entire sample.

As discussed in section \ref{dusty}, the dust release in Sextans A is mostly
provided by $\sim$ twenty AGB sources, located on the red side of the observational color-magnitude  
diagrams shown in Figs.~\ref{fcmd1} and \ref{fmiri}, whose SEDs exhibit a strong 
IR excess. This result is similar to those obtained by \citet{matsuura09, boyer11, boyer12, 
riebel12, sundar16}, who found that about $90-95\%$ of the dust in the 
SMC and LMC is coming from the reddest, ``extreme'' AGB stars. 
The discussion in section \ref{dusty} outlined that the $\sim$ twenty sources considered
here are carbon stars, which leads to the conclusion that the global DPR of
Sextans A is dominated by this class of objects. 
The dominant role of C-stars in the dust production can be explained
based on the arguments presented in section \ref{tracks}, which focus on the
sensitivity of the mineralogy of the dust released
by AGB stars to the progenitor mass and to the evolutionary phase. 

Little or no dust is released by low-mass stars that fail to reach the C-star 
stage, i.e., those discussed in point a) in section \ref{tracks}: indeed the 
DPR is generally low during the O-rich phases of low-mass stars
\citep{ventura12}, and even more so in the case of Sextans A, whose stellar 
population is metal-poor.
Significant dust production takes place in the winds of massive AGBs 
during the evolutionary phases near the luminosity peak, when the hot 
bottom burning intensity reaches its maximum \citep{ventura12, ventura14}. 
The contribution from this class of objects, discussed in point c) 
in section \ref{tracks}, is significant in galaxies hosting a solar or 
slightly sub-solar metallicity young stellar component, such as the 
LMC \citep{flavia15b}, M33 \citep{javadi13}, and M31 \citep{cla25}. 
Although some studies (e.g, \citet{boyer15b}) indicate substantial 
dust production from AGB stars even at low metallicity, in the 
specific case of Sextans A the relative contribution of refractory 
dust species from O-rich and massive AGBs is expected to be modest, 
because of the scarcity of silicon, aluminium and iron. In contrast, 
carbon dust production remains efficient, as discussed above.
Overall, even considering that all the sources populating
the $\rm (F277W-F444W) < 0.2$ mag region of the $\rm (F277W-F444W, F444W)$ diagram shown in
Fig.~\ref{fcmd1} are M-type stars, their individual DPRs barely
reach $\rm 10^{-10}~M_{\odot}/$yr (see Fig.~\ref{fcoldpr}), thus indicating that their
contribution to the global DPR of the galaxy is below $1\%$.

Given the poor contribution of low-mass stars and massive AGBs, 
most of the dust release comes from the stars that become C-stars
(point b) in section \ref{tracks}),
descending from progenitors of mass in the $\rm 1-3~M_{\odot}$ range, formed between 0.5 and 6-7 Gyr ago. For the same reasons related
to the low metallicity of Sextans A stars, we can restrict our attention
to the C-rich phase, neglecting the dust formed during the initial part
of the AGB evolution, during which the stars are O-rich.

The evolution of these stars, in terms of change in the surface
chemical composition, rate of dust release and change in the
$\rm (F277W-F444W)$ color, is shown in Fig.~\ref{fevol}. We note the significant
rise in the DPR during the C-rich phases, which increases as more carbon is carried to the surface regions by the various third dredge-up
episodes. This leads to redder $\rm (F277W-F444W)$ colors, owing to
the effects of dust reprocessing on the shape of the SED. During the C-star phase, the DPR is correlated with $\rm (F277W-F444W)$, as shown in Fig.~\ref{fcoldpr}. 
It increases from $\rm 10^{-8}~M_{\odot}/$yr at $\rm (F277W-F444W)=1$ mag,
to $\rm 10^{-7}~M_{\odot}/$yr at $\rm (F277W-F444W)=4$ mag, up to
$\rm 10^{-6}~M_{\odot}/$yr at $\rm (F277W-F444W)=8$ mag. The results 
reported in Fig.~\ref{fcoldpr} indicate an overall spread in the 
DPR at a given $\rm (F277W-F444W)$ of a factor $\sim 2$, due to the differences among the DPRs of stars of 
different mass evolving at the same $\rm (F277W-F444W)$ color, in turn
related to the differences in the progenitor masses, in the
$\rm 1-3~M_{\odot}$ range. Indeed the stars descending from higher
mass progenitors evolve at higher AGB luminosities, which favor
larger mass loss rates, and consequently a higher efficiency of the dust
formation process.
 
The predicted tight relation between the DPR and the $\rm (F277W-F444W)$ color shown in Fig.~\ref{fcoldpr} is a promising diagnostic for estimating dust production by AGB stars using JWST/NIRCam data. In the $\rm (F277W-F444W) > 0.5$ mag domain this relation can be
approximated by:

$$
\rm \frac{\dot M_{dust}}{10^{-8} \frac{M_{\odot}}{yr}} = (F277W-F444W)^{1.7} .
$$

The results from AGB evolution and dust formation modelling indicate that most of the dust is produced by stars with $\rm (F277W-F444W) > 0.5$ mag. Using the relation above, 
we find that the reddest stars in Sextans A have DPRs ranging between $10^{-10}$ and $\rm 10^{-7}~M_{\odot}/$yr. These stars are expected to produce primarily amorphous carbon dust, with a minor contribution from SiC (below $5\%$) in terms of the production rate. The bluer stars, while more numerous, contribute a negligible amount to the total dust production.
These findings are consistent with the common understanding of dust production
in metal-poor environments, such as Sextans A, where in fact we expect a dominant
contribution from solid carbon to the global dust budget.

The conclusions above in regard to the dustiest
objects of Sextans A are based on the assumption that they
are not part of binary systems. Leaving aside the unlikely possibility
that these objects are currently
experiencing a strong binary interaction, and the irrelevant (for the
present analysis) case that they belong to binary systems with large 
separation, such that no mass transfer took place, we consider the 
possibility that these sources are the former secondary components of 
binary systems and received mass from the companion, when the latter was 
evolving through the RGB. The mass transfer would result in higher 
envelope masses at the start of the AGB phase, so that their evolution 
would be similar to that experienced by higher-mass progenitors. The 
results in terms of dust production given above, particularly the relation
between DPR and the $\rm (F277W-F444W)$ color, would not be significantly
affected; with regard to the characterization of these sources, 
discussed in section \ref{dusty}, the derivation of the mass
at the start of the core helium-burning phase $\rm M_{HB}$ would 
remain substantially unchanged, while the progenitor masses 
$\rm M_{ZAMS}$ would be smaller and the derived ages correspondingly older.

\section{Conclusions}
\label{concl}
We use recent JWST observations to study the evolved stellar population 
of the metal-poor galaxy Sextans A. The proposed interpretation is based 
on the comparison between the distribution of the stars in the various 
observational color-magnitude diagrams, built by combining NIRCam and MIRI data, with the 
evolutionary tracks of stars of different mass and metallicity 
$\rm Z=0.001$ (corresponding to $\rm [Fe/H] \sim -1.4$),
obtained by means of stellar evolution and dust formation modelling.

Over 90$\%$ of the AGB sources are distributed along an approximately 
vertical sequence at $\rm (F277W-F444W) \sim 0$ mag, on the blue side of the 
$\rm (F277W-F444W, F444W)$ diagram, and exhibit little or no IR excess, 
indicating negligible ongoing dust production. This region of the diagram
harbors a miscellany of stars, ranging from old, low-mass stars,
now experiencing the late phases of their AGB lifetime, to the progeny
of $\rm 1-3~M_{\odot}$ stars, taken during the evolutionary phases preceding
the formation of large quantities of carbonaceous dust, and young, massive
AGBs, now evolving through the early-AGB or the initial part of the AGB phase, 
before the activation of strong hot bottom burning conditions. The majority of the stars 
in the vertical strip are interpreted as M-type objects, although we also 
expect the presence of a small fraction of carbon stars, evolving in the 
17.4 mag $< \rm F444W < 18.2$ mag region. For all these sources we find a 
tight correlation between the $\rm F444W$ flux and luminosity, which makes
the measured $\rm F444W$ a robust luminosity indicator.

The evolved stellar population of Sextans A includes a subset of $\sim$ twenty stars 
populating the red side of the various observational color-magnitude diagrams considered, which 
exhibit significant IR emission. As far as the $\rm (F277W-F444W, F444W)$ diagram is
concerned, the position of these red objects is well reproduced by 
the evolutionary tracks of $\rm 1.25-2.5~M_{\odot}$ stars during the final part
of the AGB evolution, after significant quantities of carbon were accumulated
in the surface regions and dust was produced at high rates. 
We find that the majority of these objects, including five of the six stars
with MIRI-LRS spectroscopy recently studied by \citet{boyer25}, descend from $\rm 1.25-1.5~M_{\odot}$ 
progenitors, formed 2-3 Gyr ago. These stars are currently producing dust at rates 
between $10^{-8}$ and $\rm 10^{-7}~M_{\odot}/$yr, with amorphous carbon being the 
dominant component and SiC contributing less than 5\%.

These results are further confirmed by the analysis of the distribution of
the sources in the $\rm (F444W-F1000W, F444W)$ and $\rm (F444W-F1130W, F444W)$
diagrams. In the latter cases, the masses derived for the progenitor stars are
slightly higher than those estimated from the NIRCam data alone:
this is likely connected to an overestimation of the SiC production, possibly
suggesting that the oldest stellar population of the galaxy is more metal-poor, 
with $\rm [Fe/H] \sim -1.8$.

Overall, this study confirms the key role of low- and 
intermediate-mass AGB stars in dust enrichment, especially in low-metallicity environments, 
and offers critical constraints for refining theoretical models of dust production. The combination of NIRCAM and MIRI data allows the determination of the quantity and mineralogy of the dust produced by the individual sources, which turns of paramount importance for a critical evaluation of the predictions from dust formation modelling in the winds of evolved stars.

\begin{acknowledgements}
This work is based on observations made with the NASA/ ESA/CSA James Webb Space Telescope. The data were obtained from the Mikulski Archive for Space Telescopes at the Space Telescope Science Institute, which is operated by the Association of Universities for Research in Astronomy, Inc., under NASA contract NAS 5-03127 for JWST. These observations are associated with program JWST-GO-1619.\\
D.A.G.H. acknowledges the support from the State Research Agency (AEI) of the Ministry of Science, Innovation and Universities (MICIU) of the Government of Spain, and the European Regional Development Fund (ERDF), under grant PID2023-147325NB-I00/AEI/10.13039/501100011033. This publication is based upon work from COST Action CA21126 - Carbon molecular nanostructures in space (NanoSpace), supported by COST (European Cooperation in Science and Technology).\\
RDG was support, in part, by the United States Air Force. 
FK acknowledges support from the Spanish Ministry of Science, Innovation and Uni
versities, under project PID2023-149918NB-I00, financed by MCIU /AEI /10.13039/5
01100011033 / FEDER, EU.\\
This work was also partly supported by the Spanish program Unidad de Excelencia 
María de Maeztu CEX2020-001058-M, financed by MCIN/AEI/10.13039/501100011033.\\
RS’s contribution to the research described here was carried out at the Jet Propulsion Laboratory, California Institute of Technology, under a contract with NASA (80NM0018D0004), and partially funded by award JWST-GO-01619.003-A from the STScI under NASA contract NAS5-03127.
\end{acknowledgements}

\facilities{JWST(NIRCam, MIRI)}

\software{{\sc dolphot} \citep{dolphin00, dolphin16}, JWST Pipeline \citep{pipeline}
}

%
%


\end{document}